\documentclass[aps,reprint,showpacs,twocolumn,superscriptaddress,floatfix]{revtex4-1}
\usepackage{epsfig}
\usepackage[T1]{fontenc}
\usepackage[utf8]{inputenc}
\usepackage{amsmath}
\usepackage{amssymb}
\usepackage{amsfonts}
\usepackage{mathptmx}
\usepackage{textcomp}
\usepackage{dcolumn}
\usepackage{eucal}
\usepackage{bm}
\usepackage{color}
\usepackage[colorlinks,linkcolor=blue,citecolor=blue]{hyperref}

\usepackage{epstopdf}

\begin{document}


\title{Towards phase-coherent caloritronics in superconducting circuits}

\author{Antonio Fornieri}
\email{antonio.fornieri@sns.it}
\affiliation{NEST, Istituto Nanoscienze-CNR and Scuola Normale Superiore, Piazza S. Silvestro 12, I-56127 Pisa, Italy}

\author{Francesco Giazotto}
\email{francesco.giazotto@sns.it}
\affiliation{NEST, Istituto Nanoscienze-CNR and Scuola Normale Superiore, Piazza S. Silvestro 12, I-56127 Pisa, Italy}


\date{\today}


\begin{abstract}
\textbf{The emerging field of phase-coherent caloritronics (from the Latin word "calor", i.e., heat) is based on the possibility to control heat currents using the phase difference of the superconducting order parameter. The goal is to design and implement thermal devices able to master energy transfer with a degree of accuracy approaching the one reached for charge transport by contemporary electronic components. This can be obtained by exploiting the macroscopic quantum coherence intrinsic to superconducting condensates, which manifests itself through the Josephson and the proximity effect. Here, we review recent experimental results obtained in the realization of heat interferometers and thermal rectifiers, and discuss a few proposals for exotic non-linear phase-coherent caloritronic devices, such as thermal transistors, solid-state memories, phase-coherent heat splitters, microwave refrigerators, thermal engines and heat valves. Besides being very attractive from the fundamental physics point of view, these systems are expected to have a vast impact on many cryogenic microcircuits requiring energy management, and possibly lay the first stone for the foundation of electronic thermal logic.} 
\end{abstract}


\maketitle

In the last decades, the impressive evolution of modern electronics has reached a point where quantum effects and phase coherence are ordinarily exploited to study exotic phenomena at the nanoscale under controlled and adjustable conditions. Only very recently, instead, scientists have started to exploit the great potentialities offered by nanotechnology for the investigation and control of heat currents (the branch of science called "caloritronics"). Interesting advances in the understanding of fundamental properties of thermal transport have been obtained in experiments involving atomic or molecular junctions~\cite{DubiRev}. A few works~\cite{Schwab,Meschke,Jezouin} have shown that heat flow has a quantum limit - just as the electric current - and that this limit does not depend on the nature of heat carriers. Furthermore, a remarkable amount of theoretical and experimental studies has been focused on the interaction between heat and spin currents in thermoelectric devices~\cite{BauerRev}. 

From the point of view of applications, the largest effort has been put into electronic and phononic thermometry or refrigeration~\cite{GiazottoRev,MuhonenRev,MottonenArx}, but the most intriguing and ambitious goal has always been the full control of heat currents, aiming to emulate the accuracy regularly obtained for charge transport in modern electronic devices. This attracting possibility was first envisioned by a conspicuous amount of theoretical works designing non-linear phononic devices~\cite{LiRev}. However, the practical realization of these structures still appears challenging, hampering significant improvements of the first promising results~\cite{Zettl,Tian}. An appealing alternative is represented by the notable ingredient of phase coherence, whose role in thermal transport was almost unknown until ten years ago, with just the exceptions of Refs.~\citenum{Chandrasekhar1,Chandrasekhar2,Vinokur}. This gap was filled by the birth of \textit{phase-coherent caloritronics}~\cite{Meschke,GiazottoNature,MartinezRev}, a young field of nanoscience that takes advantage of the long-range phase coherence of the superconducting condensate to manipulate electronic and photonic heat currents in solid-state mesoscopic circuits. As we shall argue, superconducting phase coherence represents a unique control knob for heat flows and allows to design several non-linear caloritronic devices to obtain an unprecedented thermal management at the nanoscale.

\begin{figure*}
\centering
\includegraphics[width=0.7\textwidth]{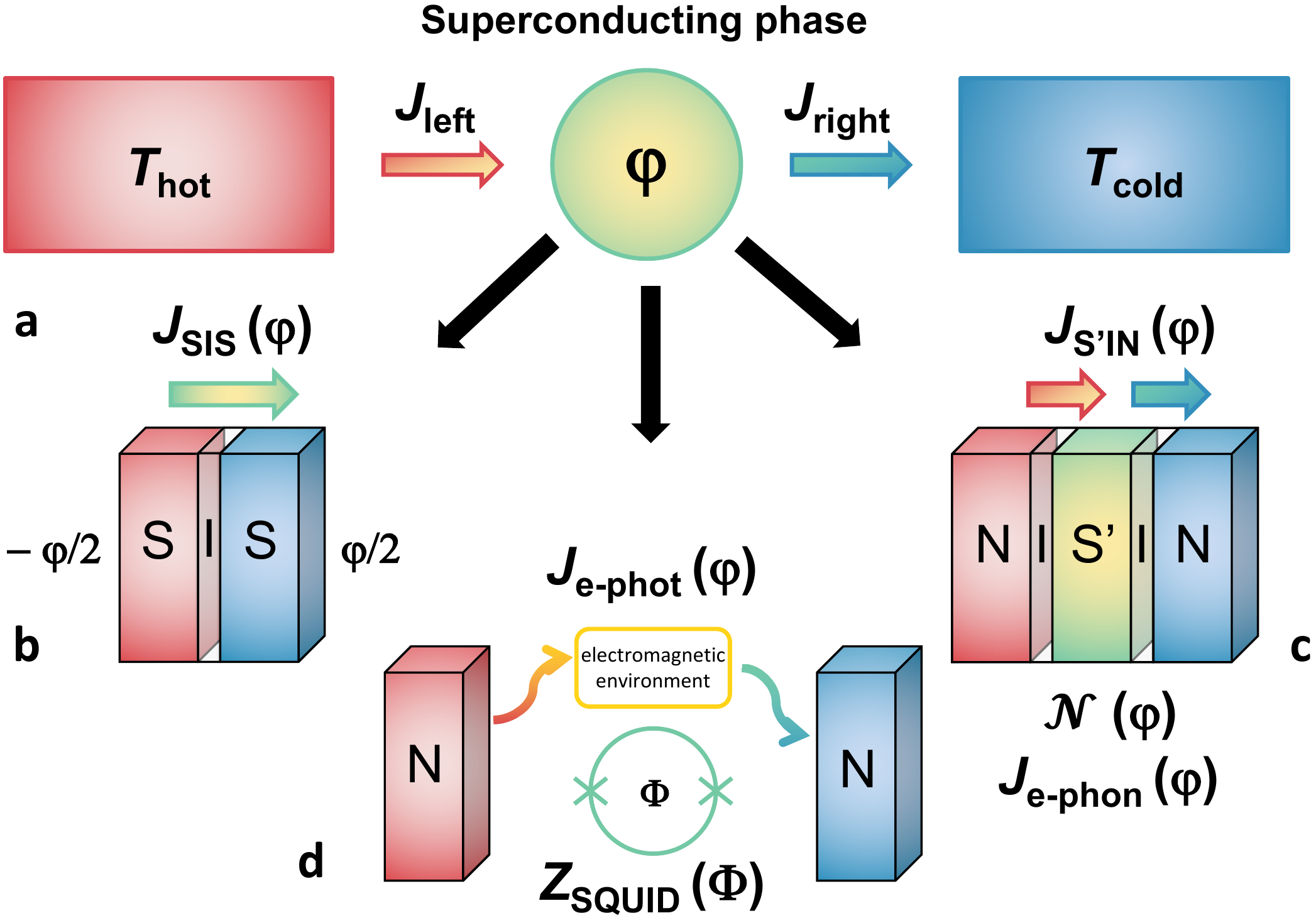}
\caption{\textbf{Physical picture at the basis of phase-coherent caloritronics.} \textbf{a.} Scheme of the fundamental idea: two electronic thermal reservoirs residing at different temperatures ($T_{\rm hot}>T_{\rm cold}$) can exchange energy by means of a physical mechanism controlled by the superconducting phase difference $\varphi$. The latter is used as a knob to control the electronic heat currents $J_{\rm left}$ and $J_{\rm right}$ flowing from the hotter to the colder reservoir. This principle can be implemented through the following approaches. \textbf{b.} The first one exploits the Josephson effect: the electronic heat current $J_{\rm SIS}(\varphi)$ flowing through a Josephson tunnel junction depends on the macroscopic phase difference between the superconducting condensates. \textbf{c.} The second approach consists in the phase manipulation of the density of states $\mathcal{N}(\varphi)$ in a superconducting proximity layer S' by means of a SQUIPT (see text). This possibility enables the phase control of thermal currents exchanged with the reservoirs $J_{\rm S'IN}(\varphi)$ and with the lattice phonons $J_{\rm e-phon}(\varphi)$. \textbf{d.} Last approach relies on the electron-photon coupling to exchange energy between two electronic reservoirs not in galvanic contact. The resulting heat current $J_{\rm e-phot}(\varphi)$ can be regulated via an intermediate phase circuit, such as DC SQUID with a magnetic-flux-dependent impedance $Z_{\rm SQUID}(\Phi)$. In all panels S stands for superconductor, N for normal metal and I for insulator.
\label{Fig1}}
\end{figure*}

The physical picture at the basis of phase-coherent caloritronics is represented schematically in Fig.~\ref{Fig1}a. The fundamental idea is to exploit a suitable physical effect that depends on the superconducting phase difference $\varphi$ to control the electronic heat flow ($J_{\rm left}$ and $J_{\rm right}$) between two electronic thermal reservoirs residing at temperatures $T_{\rm hot}>T_{\rm cold}$. Towards this end, three main different approaches can be followed. As shown in Fig.~\ref{Fig1}b, the first possibility consists in the phase control of heat transported by electrons in a temperature-biased Josephson tunnel junction (JJ), since the electronic thermal current $J_{\rm SIS}$ flowing through the junction depends on the macroscopic phase difference between the superconducting electrodes S~\cite{MakiGriffin,GiazottoAPL,GiazottoNature}. Although this method has been the most prolific from the experimental point of view~\cite{GiazottoNature,MartinezNature,MartinezNatRect,FornieriNature,FornieriArxiv}, it allows to only partially modulate $J_{\rm SIS}$, as it will be discussed in the following.

The second approach, instead, is sketched in Fig.~\ref{Fig1}c and relies on the phase manipulation of the electronic thermal conductivity and the electron-phonon coupling in phase-engineered superconducting proximity systems. Here, two normal metal (N) reservoirs are tunnel-coupled to an other normal layer (S') in which superconducting correlations are induced thanks to the proximity effect. As we shall argue, S' can be inserted in a superconducting loop forming a superconducting quantum interference proximity transistor (SQUIPT)~\cite{SQUIPT1,SQUIPT2,SQUIPT3}, which can be used to tune the phase difference across the proximized layer and therefore its density of states (DOS) $\mathcal{N}(\varphi)$~\cite{Zhou,leSueur}. The latter directly affects the thermal properties of S', leading to the phase manipulation of the electronic heat currents exchanged with the reservoirs $J_{\rm S'IN}(\varphi)$ and with the lattice phonons $J_{\rm e-phon}(\varphi)$. Even though an experimental proof is still lacking, this method could in principle provide variations of the thermal conductance of several orders of magnitude~\cite{StrambiniAPL}.

Last approach is sketched in Fig.~\ref{Fig1}d and is based on the regulation of the energy exchanged between electrons and photons thanks to the phase tuning of the coupling with the electromagnetic environment. In contrast to the previous cases, in general the reservoirs are \textit{not} in galvanic contact and the photonic heat current $J_{\rm e-phot}(\varphi)$ is controlled by an intermediate superconducting circuit [for instance, a direct current superconducting quantum interference device (DC SQUID)] inductively or capacitively coupled to the reservoirs~\cite{Ojanen,Pascal}. The SQUID is characterized by a magnetic-flux-dependent impedance $Z_{\rm SQUID}(\Phi)$ that can act as a \textit{contactless} knob for $J_{\rm e-phot}(\varphi)$. This method has been already demonstrated experimentally~\cite{Meschke}, but its potential is far from being fully exploited.

All the options listed above can be implemented in the framework of quasiequilibrium regime~\cite{GiazottoRev}: since the electron-phonon coupling in metals is strongly suppressed at temperatures below 1 K, one can inject a Joule power in the structure, thereby driving the electronic system into a Fermi distribution function characterized by a temperature $T_{\rm e}$ that can be significantly different from that of the lattice phonons. In our case, the latter are fully thermalized with the substrate phonons residing at the bath temperature $T_{\rm bath}$, thanks to the vanishing Kapitza resistance between thin metallic films and the substrate~\cite{GiazottoNature,Wellstood,MartinezNature,MartinezNatRect,FornieriNature,FornieriArxiv}. Typically, the superconducting parts are implemented by aluminum (Al) electrodes, which are well known to form high-quality tunnel junctions. At low temperatures, the heat current $J_{\rm e-phon}$ released by the electrons to the phonon bath is exponentially suppressed by the superconducting energy gap~\cite{Timofeev1}.
On the other hand, N electrodes are usually made of copper (Cu) or manganese-doped aluminum (Al$_{0.98}$Mn$_{0.02}$)~\cite{Meschke,GiazottoNature,MartinezNature,MartinezNatRect,FornieriNature,FornieriArxiv}. The former material is particularly suited to be coupled to Al leads in order to form superconducting proximity structures~\cite{SQUIPT1,SQUIPT2,SQUIPT3} and is characterized by $J_{\rm e-phon}\sim (T_{\rm e}^5-T_{\rm bath}^5)$~\cite{GiazottoRev,Wellstood}. In Al$_{0.98}$Mn$_{0.02}$, instead, $J_{\rm e-phon}\sim (T_{\rm e}^6-T_{\rm bath}^6)$~\cite{Maasilta}, thus reducing phononic losses at low temperatures. Moreover, its good oxidation properties make Al$_{0.98}$Mn$_{0.02}$ very useful to realize NIN and NIS structures~\cite{MartinezNature,MartinezNatRect,FornieriNature}.

In the following sections, we shall analyze more deeply each different approach, reviewing the major experimental achievements and the existing proposals for novel caloritronic devices.

\section*{Josephson tunnel circuits}

\begin{figure*}[t]
\centering
\includegraphics[width=1\textwidth]{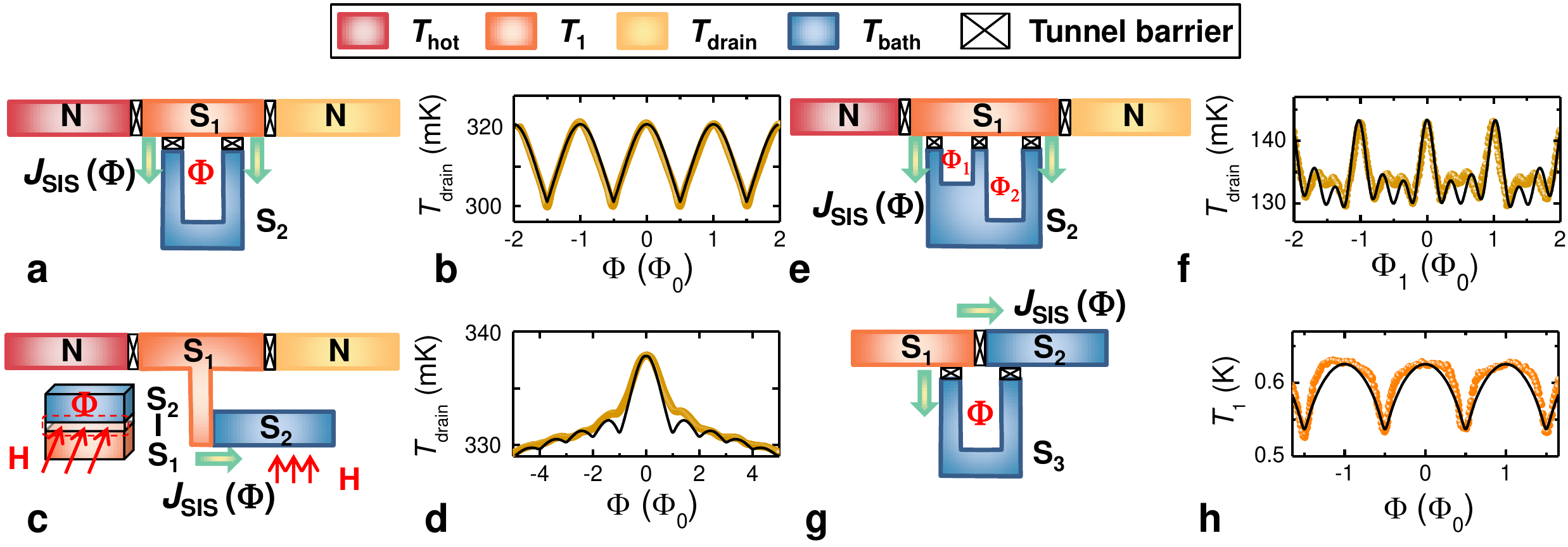}
\caption{\textbf{Josephson heat interferometers.} \textbf{a.} Schematic representation of the thermal counterpart of a DC SQUID. The interferometer is composed by two superconducting electrodes S$_1$ and S$_2$ residing at temperatures $T_1$ and $T_{\rm bath}$, respectively, and forming a loop with two JJs in parallel. The upper branch S$_1$ is heated by the N source (residing at $T_{\rm hot}$) and the resulting thermal gradient across the SQUID generates a tunable heat current $J_{\rm SIS}$. \textbf{b.} Drain temperature $T_{\rm drain}$ modulations as a function of the magnetic flux $\Phi$ threading the loop of the DC SQUID for $T_{\rm hot}=675$ mK and $T_{\rm bath}=235$ mK. \textbf{c.} Schematics of a thermal diffractor. The core of the interferometer consists of an extended rectangular JJ pierced by a flux $\Phi$ controlled by an in-plane magnetic field $H$. \textbf{d.} Fraunhofer-like interference pattern of $T_{\rm drain}$ as a function of $\Phi$ for $T_{\rm hot}=720$ mK and $T_{\rm bath}=240$ mK. \textbf{e.} Sketch of a Josephson heat modulator. The central part of the structure consists of a double-loop SQUID threaded by the magnetic fluxes $\Phi_1$ and $\Phi_2$, with $\Phi_2\simeq 2 \Phi_1$. \textbf{f.} Interference pattern of $T_{\rm drain}$ for $T_{\rm hot}=560$ mK and $T_{\rm bath}=25$ mK. \textbf{g.} Scheme of a "pseudo" radio-frequency SQUID able to polarize the JJ between S$_1$ and S$_2$ from $\varphi=0$ to $\varphi=\pi$, thereby realizing a $0-\pi$ controllable thermal JJ. \textbf{h.} Electronic temperature $T_1$ of the electrode S$_1$ vs. $\Phi$ for an injected Joule power of $112$ pW at $T_{\rm bath}=25$ mK. In all schematics we assumed $T_{\rm hot}>T_1>T_{\rm drain}>T_{\rm bath}$, while the green arrows represent the direction of the flux-controlled heat current $J_{\rm SIS}$. In all graphs, circles represent the experimental data, whereas solid black lines are the fits obtained from the thermal model of each structure (see text). Data are taken from Refs~\citenum{GiazottoNature,MartinezNature,FornieriNature,FornieriArxiv}.  
\label{Fig2}}
\end{figure*}

Just three years after the prediction of the Josephson effect~\cite{Josephson}, Maki and Griffin~\cite{MakiGriffin} calculated the expression accounting for the electronic heat current flowing through a temperature-biased SIS JJ (where S stands for superconductor and I for insulator)~\cite{Guttman,GiazottoAPL,Zhao}:
\begin{equation}
J_{\rm SIS}(\varphi)=J_{\rm qp}-J_{\rm int}\rm\; cos \varphi.\label{Jtot}
\end{equation} 
Equation~\ref{Jtot} contains interesting information about the impact of the Josephson effect on thermal transport. First of all, $J_{\rm SIS}$ consists of two components: the former accounts for the heat carried by quasiparticles and represents an incoherent flow of energy from the hot to the cold reservoir~\cite{GiazottoRev,Tinkham}. On the other hand, $J_{\rm int}$ is the thermal counterpart of the "quasiparticle-pair interference" term that contributes also to the \textit{electrical} current tunneling through a JJ~\cite{Barone,Pop}. It stems from energy-carrying tunneling processes involving concomitant creation and destruction of Cooper pairs on both sides of the junctions~\cite{Barone,Guttman} and is therefore regulated by the phase difference $\varphi$ between the two superconducting condensates. 
Depending on $\varphi$, this phase-coherent component can flow in the opposite direction with respect to that imposed by the temperature gradient, although the total $J_{\rm SIS}$ follows the second principle of thermodynamics, i.e., $J_{\rm qp}$ is always greater than $J_{\rm int}$. As mentioned in the introduction, this inequality represents the main limit of this approach based on Josephson tunnel junctions, which can only partially modulate electronic heat currents.

All these features have been investigated and confirmed experimentally in different interferometer-like structures, which are schematically depicted in Fig.~\ref{Fig2}. The first three devices have a similar structure: two normal metal electrodes N are used as electronic thermal reservoirs connected by a superconducting central part that forms the core of the interferometer. The reservoirs are also connected to superconducting wires that create SINIS junctions, which can be used as Joule heaters or thermometers (not shown)~\cite{GiazottoRev}. By injecting a Joule power, one can heat the electrons in the source up to a temperature $T_{\rm hot}$, so as to elevate the quasiparticle temperature $T_1$ of the upper branch S$_1$ of the interferometers above $T_{\rm bath}$. On the contrary, the lower branch S$_2$ is thermally anchored to the bath temperature, thanks to its large volume. In this way, it is possible to obtain a substantial thermal gradient between S$_1$ and S$_2$, thereby generating finite heat currents $J_{\rm SIS}$ flowing through the JJs that form the central part of the structure. By applying an external magnetic flux $\Phi$ one can tune the phase polarization of the JJs~\cite{Tinkham} and therefore manipulate $J_{\rm int}$, leading to phase-coherent oscillations of the drain temperature $T_{\rm drain}$.

\begin{figure*}
\centering
\includegraphics[width=0.7\textwidth]{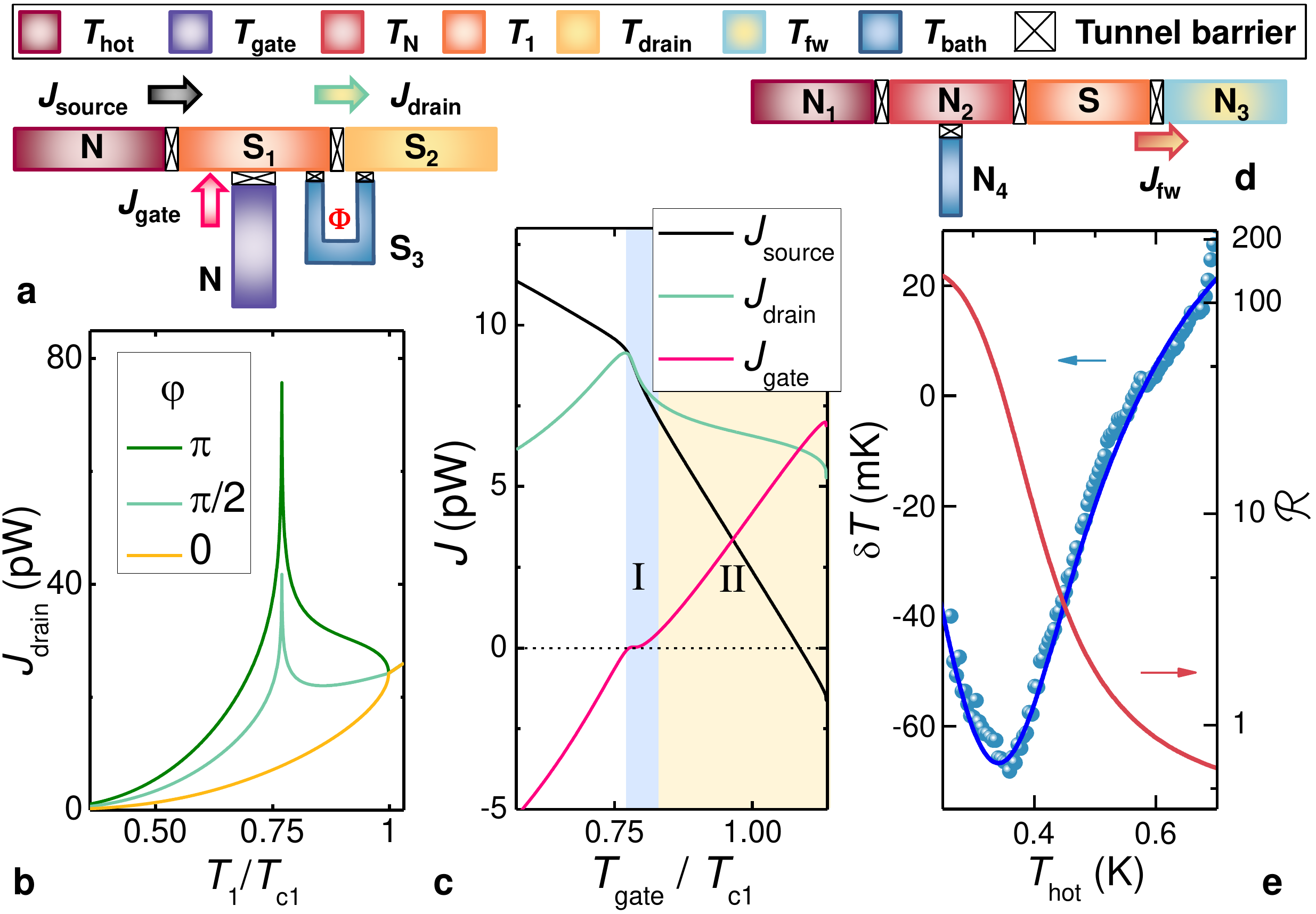}
\caption{\textbf{Thermal transistors and thermal rectifiers.} \textbf{a.} Schematic representation of a possible thermal transistor acting as a heat modulator, with $T_{\rm hot}\sim T_{\rm gate}>T_1>T_{\rm drain}$. The phase difference $\varphi$ across the S$_1$IS$_2$ JJ can be tuned form 0 to $\pi$ thanks to a "pseudo" rf SQUID as the one presented in Figs.~\ref{Fig2}g and~\ref{Fig2}h. \textbf{b.} Calculated electronic heat current $J_{\rm drain}\equiv J_{\rm SIS}$ flowing through the JJ as a function of T$_1$ for different values of $\varphi$. The curves are obtained for $T_{\rm drain}=0.01 T_{\rm c1}$ (being $T_{\rm c1}$ the critical temperature of S$_1$) and $\delta=\Delta_2(0)/\Delta_1(0)=0.75$, where $\Delta_{1,2}(0)$ are the zero-temperature energy gaps of S$_1$ and S$_2$. \textbf{c.} Calculated $J_{\rm source}$, $J_{\rm gate}$ and $J_{\rm drain}\equiv J_{\rm SIS}$ vs. $T_{\rm gate}$ for $T_{\rm hot}=0.96 T_{\rm c1}$, $T_{\rm drain}=0.03 T_{\rm c1}$, $\varphi=\pi$ and $\delta=0.75$. The horizontal dashed line indicates $J=0$, while the shadowed regions I and II correspond to different regimes of amplification (see text). \textbf{d.} Sketch of a hybrid thermal rectifier in the forward temperature bias configuration, i.e., for $T_{\rm hot}>T_{\rm N}>T_1>T_{\rm fw}>T_{\rm bath}$. The reverse configuration is obtained by heating the N$_3$ reservoir up to $T_{\rm hot}>T_1>T_{\rm N}>T_{\rm rev}>T_{\rm bath}$, where $T_{\rm rev}$ is the temperature of the N$_1$ reservoir. \textbf{e.} Experimental temperature difference $\delta T=T_{\rm rev}-T_{\rm fw}$ as a function of $T_{\rm hot}$ at $T_{\rm bath}=50$ mK (scatter, left axes) and corresponding rectification efficiency $\mathcal{R}$ (red line, right axes on a logarithmic scale) extracted from the theoretical fit (blue line). Data are taken from Ref.~\citenum{MartinezNatRect}.
\label{Fig3}}
\end{figure*} 

Figure~\ref{Fig2}a displays the schematic of the thermal counterpart of a symmetric DC SQUID, which offered the first experimental proof of Eq.~\ref{Jtot} (Ref.~\citenum{GiazottoNature}). By imposing a thermal gradient across the device and varying $\Phi$, the authors observed modulations of $T_{\rm drain}$ as large as 21 mK at $T_{\rm bath}=235$ mK and $\langle T_{\rm drain}\rangle \simeq 310$ mK, as shown in Fig.~\ref{Fig2}b. These modulations stem from the interference between the phase-coherent components of the heat currents flowing across the JJs of the SQUID, and can be fitted with a thermal model accounting for the predominant energy-exchange mechanisms present in the system (see black line in Fig.~\ref{Fig2}b). The model is based on the conservation of energy and imposes that in stationary conditions the sum of the incoming and outgoing thermal currents for each electrode of the structure must be equal to zero~\cite{GiazottoNature}. 
The excellent agreement between the experiment and the theory confirms the physical picture described above.

The complementary demonstration of the phase-coherent nature of $J_{\rm int}$ was obtained one year later, with the observation of a thermal diffraction pattern in an extended rectangular JJ~\cite{MartinezNature}, as depicted in Fig.~\ref{Fig2}c. The temperature-biased JJ between S$_1$ and S$_2$ is threaded by a magnetic flux controlled by an in-plane magnetic field $H$. This generates quantum diffraction for the heat current $J_{\rm SIS}$ and produces an archetypal Fraunhofer-like pattern [$\propto |\mathrm{sin}(\pi \Phi/\Phi_0)/(\pi \Phi/\Phi_0)|$, where $\Phi_0$ is the superconducting flux quantum (Ref.~\citenum{Tinkham})] for the electronic drain temperature, as shown in Fig.~\ref{Fig2}d~\cite{GiazottoDiff,MartinezNature}. In analogy to what was done in the '60s for the electrical Josephson current, this experiment confirmed unequivocally the thermal current-phase relationship expressed by Eq.~\ref{Jtot}.

Although these interferometers embody the simplest structures to obtain the experimental demonstration of the prediction by Maki and Griffin, they are not ideal systems to achieve the full control of $J_{\rm int}$. Indeed, unwanted structural asymmetries (for instance, a difference in the normal-state resistance of the JJs) can reduce severely the visibility of the oscillations of a conventional single-loop SQUID~\cite{MartinezDLoop}. To avoid this, one can replace one of the JJs with an additional SQUID, thereby realizing the Josephson thermal modulator sketched in Fig.~\ref{Fig2}e~\cite{FornieriNature}, in which in principle two magnetic fluxes $\Phi_1$ and $\Phi_2$ can be driven independently. 
This device enables the generation of exotic thermal interference patterns, which are characterized by large oscillation amplitudes, high sensitivities to magnetic flux variations, and a 99 \% modulation of $J_{\rm int}$ despite the presence of a non-negligible junction asymmetry~\cite{FornieriNature}. Figure \ref{Fig2}f shows a detail around $\Phi=0$ of the temperature oscillations resulting from the double loop geometry.
Foremost, this system showed a perfect correspondence in the phase engineering of charge and thermal transport~\cite{FornieriNature}, opening the way for the conception of more sophisticated phase-coherent caloritronic devices where thermal currents can be manipulated at will.  In this perspective, the Josephson modulator could be the core of a thermal splitter~\cite{BosisioPRB,BenAbdallahAPL}, able to control the amount of energy transferred among several terminals residing at different temperatures. Or even more interesting, the magnetic fluxes threading the loops could be driven independently with the help of superconducting on-chip coils. This would allow to control separately the phase-biasing of the loops, with the possibility to perform closed cycles in the flux parameter space. Combining this possibility with the thermal rectifying properties of a JJ formed by S leads with different energy gaps~\cite{MartinezAPL} would open many opportunities to realize heat pumps~\cite{Ren}, microwave coolers~\cite{Valenzuela,SolinasPRB} or time-dependent thermal engines~\cite{Campisi,Niskanen,Quan}.

The most recent achievement in mastering $J_{\rm int}$ is represented by a $0-\pi$ phase-controllable thermal JJ. As depicted in Fig.~\ref{Fig2}g, the latter is embedded in a "pseudo" radio frequency (rf) SQUID containing three JJs, one of which supports a lower Josephson critical current with respect to the others~\cite{FornieriPRB}. This configuration enables the phase-biasing of the S$_1$IS$_2$ JJ from 0 to $\pi$ (when the flux is varied from 0 to $\Phi_0/2$), thereby allowing to minimize or maximize $J_{\rm SIS}$ and to obtain unprecedented temperature modulation amplitudes ($\sim 100$ mK at $T_{\rm bath}=25$ mK and $\langle T_{\rm 1}\rangle \simeq 570$ mK, as shown in Fig.~\ref{Fig2}h) and sensitivities to the magnetic flux exceeding 1 K$/\Phi_0$~\cite{FornieriArxiv}. The fully superconducting nature of the device allows to efficiently suppress the influence of the electron-phonon coupling, leading to a remarkably high maximum operational temperature of 800 mK. Yet, as it will be clear from the next paragraphs, this structure realizes the fundamental requirement to obtain negative differential thermal conductance (NDTC) which is at the basis of non-linear thermal devices, such as tunnel heat diodes~\cite{MartinezAPL,FornieriRev}, thermal switches and transistors~\cite{FornieriPRB}.

Figure~\ref{Fig3}a shows a possible design for a thermal transistor, in particular for a thermal modulator. The latter consists of a three-terminal device in which two N electrodes acting as source and gate are tunnel coupled to the S$_1$IS$_2$ JJ and reside at temperatures $T_{\rm hot}$ and $T_{\rm gate}$, respectively. The phase polarization of the JJ can be controlled from 0 to $\pi$ by the "pseudo" rf SQUID that we just discussed. We also define $J_{\rm source}$ as the thermal current flowing from the source to S$_1$, $J_{\rm gate}$ as the heat current flowing from the gate to S$_1$ and, lastly, $J_{\rm drain}\equiv J_{\rm SIS}$ represents the thermal flow across the JJ. Now, it is worth noting that besides phase-coherence the heat current $J_{\rm drain}$ exhibits another important feature. Figure~\ref{Fig3}b shows the calculated behavior of $J_{\rm drain}$ as a function of $T_1$ for a set drain temperature $T_{\rm drain}$ and $\delta=\Delta_2(0)/\Delta_1(0)=0.75$ [being $\Delta_{1,2}(T_{1,2})$ the temperature-dependent energy gaps of S$_1$ and S$_2$, respectively~\cite{Tinkham}]. When $\Delta_1(T_1)=\Delta_2(T_2)$, the heat current presents a sharp peak for $\varphi \neq 0$, which is due to the matching of the singularities in the superconducting densities of states~\cite{Barone}. At higher values of $T_1$ we reach the condition in which $\Delta_1<\Delta_2$ and the thermal transport across the JJ is reduced, giving rise to a region of NDTC. This effect is maximum for $\varphi=\pi$, while the peak is perfectly canceled by $J_{\rm int}$ for $\varphi=0$~\cite{FornieriPRB}. However, we can notice that the value of $J_{\rm drain}$ is always positive, confirming that $J_{\rm qp}$ is always greater than $J_{\rm int}$. A realistic system able to detect NDTC is the thermal counterpart of a tunnel diode (as the one envisioned in Ref.~\citenum{FornieriPRB}), which could also serve as a solid-state thermal memory device.
Here, we just show how NDTC in a thermal modulator can generate \textit{heat amplification}: changes in $J_{\rm gate}$ can induce a larger change in $J_{\rm source}$ and $J_{\rm drain}$, leading to a heat amplification factor $\alpha\equiv|\partial J_{\rm source,drain}/\partial J_{\rm gate}|>1$. Figure~\ref{Fig3}c shows the predicted behavior of $J_{\rm source}$ and $J_{\rm drain}$ when $T_{\rm hot}>T_{\rm drain}$ and $T_{\rm gate}$ is varied as a control knob. It is clear how in region I the device can significantly reduce both $J_{\rm source}$ and $J_{\rm drain}$ while $J_{\rm gate}$ remains close to zero and almost constant. A more quantitative analysis tells us that in this region the heat amplification factor tends to infinity, but $\alpha$ turns out to be greater than 1 also in region II, which is several hundreds of mK wide~\cite{FornieriPRB}. Alternatively, it was shown that a thermal amplifier (able to increase both $J_{\rm source}$ and $J_{\rm drain}$ as $T_{\rm gate}$ is raised above $T_{\rm drain}$) can also be designed by simply switching the position of the JJ and the N source electrode~\cite{FornieriPRB}. Even though the variations of $J_{\rm source}$ and $J_{\rm drain}$ appear to be smaller than their absolute values, these devices represent the most accessible systems in order to obtain the first experimental demonstration of heat amplification. In the last section, we will show how a more sophisticated device based on thermoelectric effect can represent a further step towards the realization of an efficient phase-coherent thermal amplifier~\cite{PaolucciAmpl}.

The last structure we wish to present in this section is a hybrid thermal rectifier, a device that allows heat to flow preferentially in one direction. Even though the system is not phase-coherent, it proves the potential of superconducting hybrid circuits as one of the best platforms to manipulate heat currents. The structure (whose schematic is shown in Fig.~\ref{Fig3}d) consists of a N$_1$IN$_2$ISIN$_3$ chain that joins two theoretical proposals~\cite{MartinezAPL,FornieriAPL}, and exhibits two different regimes of rectification. The first is based on the different temperature dependence of the DOSes of the N$_2$ and S electrodes. Indeed, the temperature dependence of the superconducting energy gap breaks the directional symmetry of thermal transport through the simple N$_2$IS junction~\cite{MartinezAPL,GiazottoRectAPL}. 
The second and most efficient mechanism relies on the asymmetric release of energy from the device to the thermal bath thanks to the N$_4$ probe acting as a cooling fin~\cite{FornieriAPL}. In the experiment, the authors imposed a temperature gradient across the device by setting the electronic temperature of N$_1$ and N$_3$ to $T_{\rm hot}$ in the forward and reverse thermal bias configuration, respectively (in Fig.~\ref{Fig3}d the device is depicted in the forward configuration). Afterwards, they measured the output temperature of the structure $T_{\rm fw}$ ($T_{\rm rev}$) of the electrode N$_3$ (N$_1$) in the forward (reverse) configuration~\cite{MartinezNatRect}. As shown in Fig.~\ref{Fig3}e, a maximum $\delta T=T_{\rm rev}-T_{\rm fw}$ exceeding 60 mK was observed at $T_{\rm bath}=50$ mK and $T_{\rm hot} \simeq 360$ mK (full circles, left axis). The analysis of the data provided a rectification ratio $\mathcal{R}=J_{\rm fw}/J_{\rm rev}$ (solid line, right axis), where $J_{\rm fw}$ and $J_{\rm rev}$ are the heat currents flowing from S to N$_3$ and from N$_2$ to N$_1$ in the two different temperature bias configurations. A significant maximum value of $\mathcal{R}\simeq 140$ was obtained, which outscored previous experimental results by more than two orders of magnitude~\cite{Zettl,Tian}. Finally it is worthwhile to emphasize that this device can be easily modified and combined with the interferometers discussed above to realize phase-coherent thermal rectifiers~\cite{MartinezAPL,FornieriRev}.

\section*{Superconducting proximity structures}

\begin{figure}
\centering
\includegraphics[width=0.98\columnwidth]{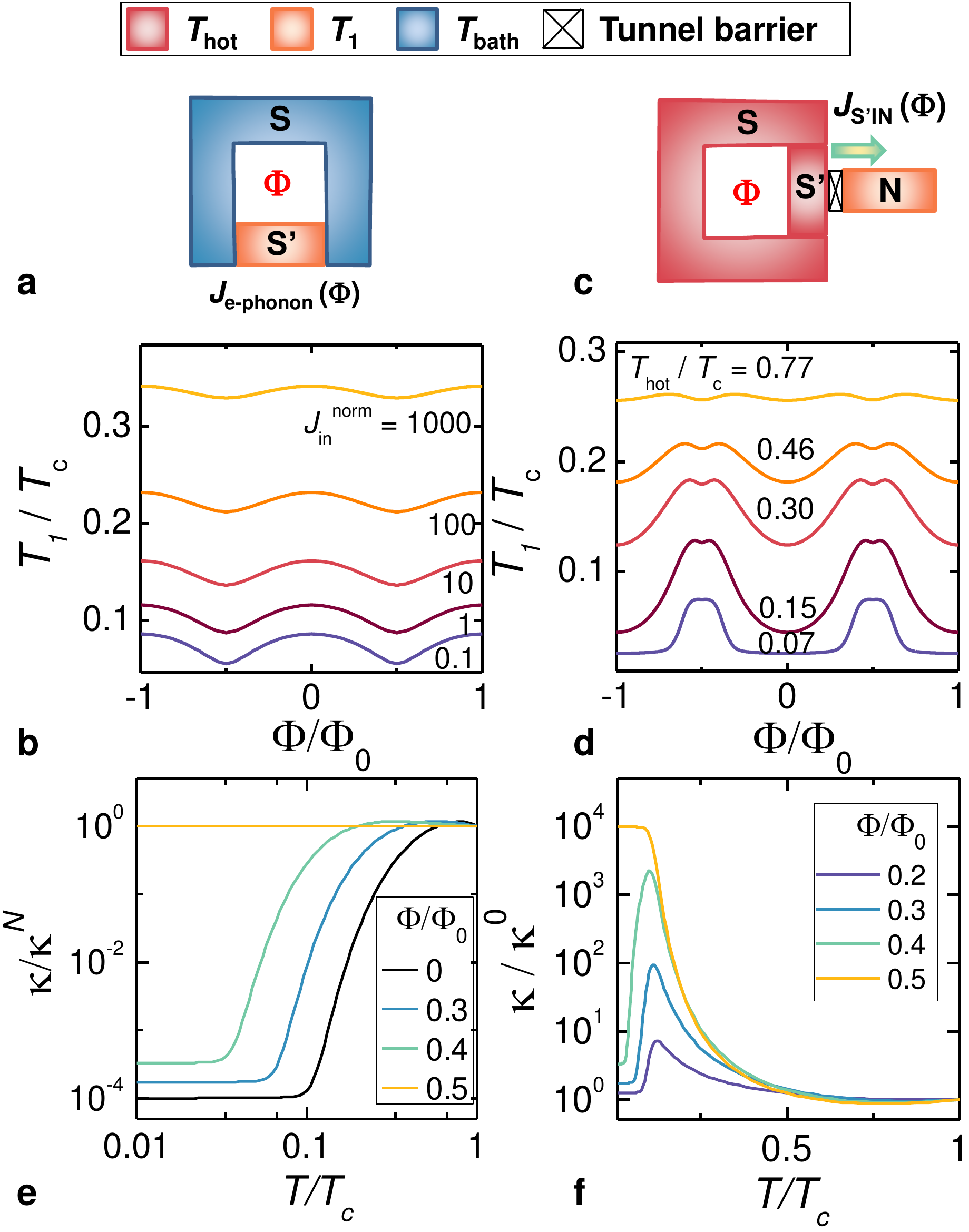}
\caption{\textbf{Superconducting proximity structures.} \textbf{a.} Scheme of a system that can be exploited to detect the phase-cherent modulation of the heat current $J_{\rm e-phon}$ due to the electron-phonon coupling in a proximity wire S' (made of Cu) inserted in a superconducting loop S (made of Al with $T_{\rm c}=1.3$ K). The electrode S' can be tunnel coupled to other superconducting electrodes that can be used as Joule heaters and thermometers (not shown)~\cite{GiazottoRev}. \textbf{b.} Calculated electronic temperature $T_1$ of the S' electrode as a function of the magnetic flux $\Phi$ threading the loop for different values of the normalized injected Joule power $J_{\rm in}^{\rm norm}$ (for details, see Ref.~\citenum{HeikkilaPRB}) at $T_{\rm bath}=50$ mK. 
\textbf{c.} Schematic representation of a proximity heat valve. The structure is similar to the one shown in panel a, but with a normal metal probe N (made of Al$_{0.98}$Mn$_{0.02}$) connected to S' by means of a tunnel junction. When S and S' are heated up to the electronic temperature $T_{\rm hot}$, the heat current $J_{\rm S'IN}$ flows across the tunnel junction S'IN and its amplitude depends on the magnetic flux $\Phi$. \textbf{d.} Calculated electronic temperature $T_1$ of the N electrode vs. $\Phi$ for different values of $T_{\rm hot}$ at $T_{\rm bath}=20$ mK. In panels a, b, c and d, we assumed $T_{\rm hot}>T_1>T_{\rm bath}$.
\textbf{e.} Normalized thermal conductance $\kappa/\kappa^{\rm N}$ of the S'IN junction vs. $T\equiv (T_{\rm hot} + T_1)/2$ calculated for $T_{\rm hot} - T_1 \ll T$ and for different values of $\Phi$. Here, $\kappa^{\rm N}$ is the normal state conductance. \textbf{f.} Normalized thermal conductance $\kappa/ \kappa^0$ as a function of $T$ for different values of $\Phi$, where $\kappa^0$ is the thermal conductance calculated for $\Phi=0$. In panels e and f, the Dynes parameter in S' is assumed to be $\Gamma=10^{-4} E_{\rm g}$~\cite{Dynes,StrambiniAPL}.\label{Fig4}} 
\end{figure} 

As mentioned in the introduction, the second approach to phase-coherent caloritronics is based on the superconducting proximity effect~\cite{DeGennes}. When a normal metal (S') is brought in contact with a superconductor, the superconducting order parameter leaks out to the normal side. The main consequences of this effect are the changes in the local DOS~\cite{Usadel,Zhou,leSueur} and the induction of a finite pair amplitude in the normal metal. For instance, in a SS'S JJ the proximity effect generates a phase-tunable minigap $E_{\rm g}(\varphi)$ in the DOS of the weak-link that has a maximum for $\varphi=0$ and vanishes for $\varphi=\pi$~\cite{leSueur,Virtanen}. This offers the opportunity to manipulate in a continuous fashion the thermal properties of the S' electrode, including its electronic entropy and specific heat~\cite{RabaniJAP,RabaniPRB}, which can be varied from those of a superconductor to those of a normal metal. Even more interesting, also the relaxation mechanisms existing in S' can be controlled by tuning the value of $E_{\rm g}$~\cite{HeikkilaPRB}. Indeed, as noted in the introduction, the presence of the energy gap in the DOS exponentially suppresses the electron-phonon coupling in a superconductor~\cite{Timofeev1,Wellstood}.

Figure~\ref{Fig4}a shows a possible configuration to detect the phase modulation of the electron-phonon coupling in a S' weak-link assumed to be made of Cu. The latter interrupts an Al loop S (with $T_{\rm c}=1.3$ K) that is used to phase-polarize the JJ by means of an external magnetic flux $\Phi$~\cite{Tinkham}. When $|\Phi|=k \Phi_0$ (being $k$ an integer), $E_{\rm g}$ is maximum and by heating S' the heat current $J_{\rm e-phon}$ released by electrons to the phonon bath is minimized. Thus, for these values of $\Phi$ the S' electronic temperature $T_1$ reaches a maximum, whereas it shows a minimum for semi-integer values of the flux quantum (i.e., for $E_{\rm g}=0$), as shown in Fig.~\ref{Fig4}b. The periodic modulations of $T_1$ can exhibit a remarkable maximum amplitude of $\sim 30$  mK at $T_{\rm bath}=50$ mK and for $\langle T_1 \rangle = 90$ mK~\cite{HeikkilaPRB}, but it is possible to envision an even more efficient way to exploit the phase-dependence of $E_{\rm g}$.

Indeed, one can think of the superconducting energy gap as a barrier for energy-carrying quasiparticles. The possibility to tune phase-coherently the height of this barrier leads promptly to the design of a valve for electronic heat currents~\cite{StrambiniAPL}. The proposed structure is shown on Fig.~\ref{Fig4}c, and consists of a SQUIPT~\cite{SQUIPT1,SQUIPT2,SQUIPT3}, i.e. an Al loop S interrupted by a normal wire S' (made of Cu), which is tunnel coupled to a normal metal probe N (made of Al$_{0.98}$Mn$_{0.02}$). If the electronic temperature of S and S' is raised up to $T_{\rm hot}$, we can obtain modulations of the N temperature $T_1$ with amplitudes exceeding 100 mK at $T_{\rm bath}= 20$ mK and $\langle T_1 \rangle = 100$ mK. 

We stress that, although not yet experimentally proven, the approach described in this section has the potential to be very effective thanks to the ability to vary the thermal conductance of the S'IN junction by several orders of magnitude, as shown in Figs.~\ref{Fig4}e and~\ref{Fig4}f. This is in contrast to previous realizations of thermal Andreev interferometers~\cite{Chandrasekhar1,Chandrasekhar2,Vinokur} and to the Josephson circuits analyzed in the previous section, which are not able to control the incoherent component of the heat current $J_{\rm qp}$.


\section*{Photonic heat transistors}

\begin{figure}
\centering
\includegraphics[width=1\columnwidth]{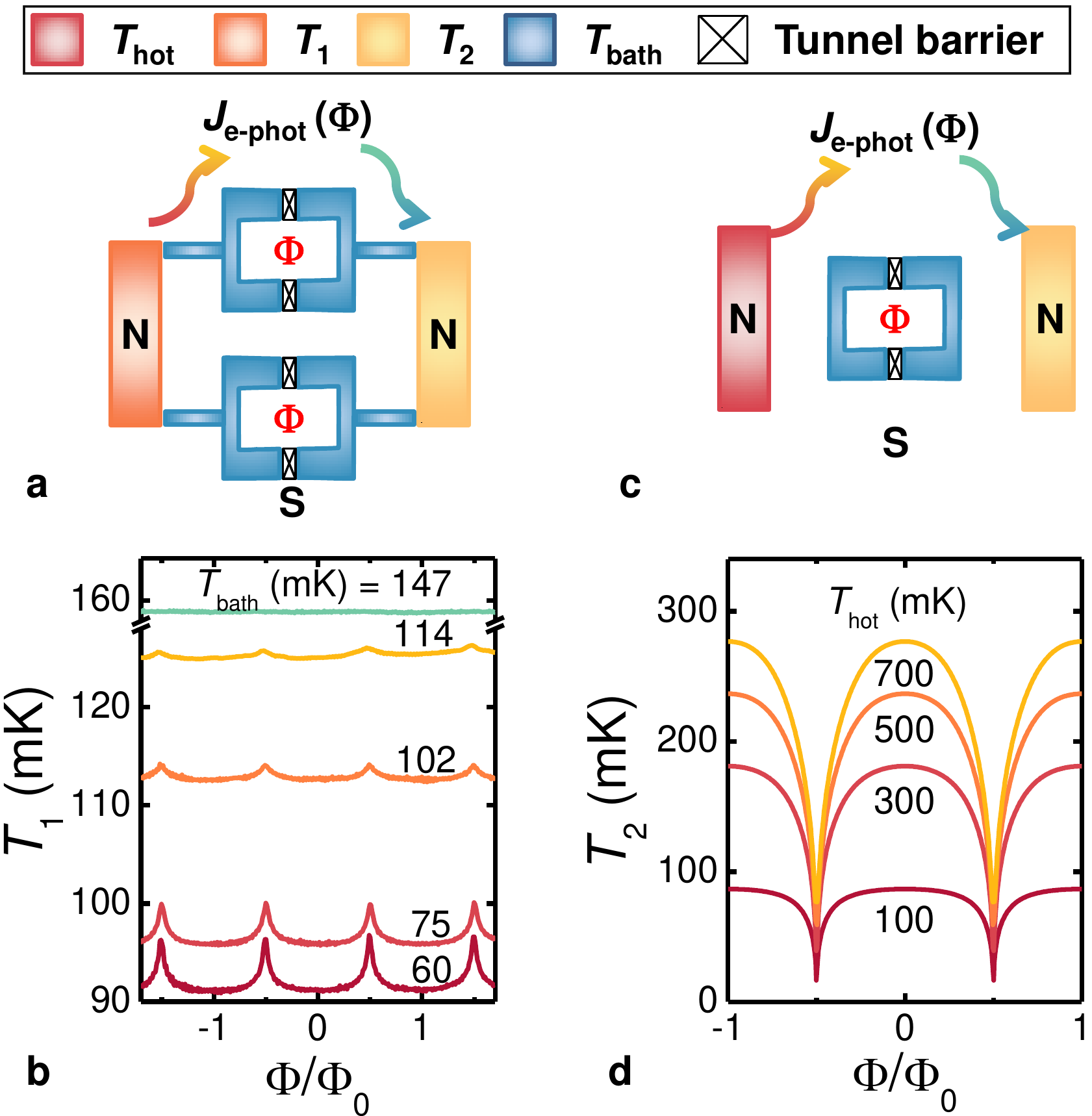}
\caption{\textbf{Photonic heat transistors.} \textbf{a.} Scheme of the first system used to demonstrate phase-coherent photonic heat conduction and its quantum limit. Two N reservoirs (made of palladium-gold) residing at $T_1>T_2>T_{\rm bath}$ are coupled by an intermediate circuit consisting of two DC SQUIDs made of Al, connected by superconducting lines \textit{in clean contact} with the reservoirs. These lines act as ideal insulators for galvanic thermal transport between the N electrodes at low temperatures, thanks to the Andreev reflection mechanism~\cite{Tinkham,Zhao} and the presence of the energy gap. The magnetic flux $\Phi$ piercing the loops can vary the Josephson inductance of the SQUIDs, thus regulating the electromagnetic coupling between the N electrodes, and modulating the photonic heat current $J_{\rm e-phot}(\Phi)$. \textbf{b.} Electronic temperature $T_1$ modulations as a function of $\Phi$ for a parasitic injected power of $1$ fW and for different values of $T_{\rm bath}$. Data are taken from Ref.~\citenum{Meschke}. \textbf{c.} Sketch of a non-galvanic photon thermal transistor. In this case, the intermediate circuit is a DC SQUID (made of Al) \textit{inductively} or \textit{capacitevily} coupled to the N reservoirs (made of Al$_{0.98}$Mn$_{0.02}$). The temperature of the right reservoir $T_2$ can be controlled by the change of the Josephson inductance when $\Phi$ is varied. \textbf{d.} Calculated magnetic interference pattern of $T_2$ for different values of $T_{\rm hot}$ at $T_{\rm bath}=10$ mK in the case of an inductive coupling. 
Here, we set a mutual inductance of $500$ pH, the capacitance of the SQUID JJs of $2$ fF and the maximum Josephson inductance of $250$ pH. 
In all the panels we assumed $T_{\rm hot}>T_1>T_2>T_{\rm bath}$.
\label{Fig5}} 
\end{figure}

Last approach to obtain phase control of heat currents is through the transport of thermal photons, which becomes the dominant mechanism when the phononic and the electronic channels are frozen~\cite{schmidt,Meschke}. Figure~\ref{Fig5}a shows the first experimental configuration that was used to detect and phase control the photonic thermal transport between two N reservoirs with finite resistances $R_1$ and $R_2$~\cite{Meschke}. When the electronic temperature of the left reservoir is raised up to $T_1>T_{\rm bath}$, the electromagnetic noise power radiated from this resistor generates a net photonic power $J_{\rm e-phot}$ flowing towards the other reservoir residing at temperature $T_2$ (with $T_1>T_2>T_{\rm bath}$). Since the dimensions of the circuit are typically much smaller than the photon thermal wavelength ($\lambda_{\rm phot}>1$ cm at temperatures below 1 K)~\cite{schmidt,Meschke}, it is possible to consider the structure as a lumped series of equivalent circuits~\cite{schmidt,Pascal,Paolucci}. It has been shown that this circuital approach is equivalent to the analysis employing nonequilibrium Green's functions~\cite{Ojanen}. It can be demonstrated that $J_{\rm e-phot}$ is proportional to a frequency-dependent transfer function~\cite{schmidt,Pascal}:
\begin{equation}
\mathcal{T}(\omega)=\frac{4 \Re [Z_1 (\omega)] \Re [Z_2 (\omega)]}{|Z_{\rm tot}(\omega)|^2},
\end{equation} 
where $Z_{1,2}(\omega)$ are the impedances of the N reservoirs, $Z_{\rm tot}(\omega)=Z_1(\omega)+Z_2(\omega)+Z_{\rm c}(\omega)$ is the total series impedance of the circuit and $Z_{\rm c}(\omega)$ is the impedance of the coupling circuit. In the case depicted in Fig.~\ref{Fig5}a, the coupling circuit consists of two DC SQUIDs connected to the N electrodes [made of palladium-gold, with $J_{\rm e-ph}\sim (T_{\rm e}^5-T_{\rm bath}^5)$] through superconducting Al lines, which act as ideal thermal insulators at low temperatures, thanks to the Andreev reflection mechanism~\cite{Tinkham,Zhao} and the presence of the energy gap. The SQUIDs are equivalent to two LC circuits with a variable Josephson inductance that depends on the magnetic flux $\Phi$ piercing the loops. Thus, also $\mathcal{T}$ and $J_{\rm e-phot}$ become flux-dependent, allowing a phase control of the photonic heat flux across the circuit. In this way, modulations of $T_1$ (up to $\sim 6$ mK of amplitude) as a function of $\Phi$ have been observed up to $T_{\rm bath}\simeq 150$ mK and for $\langle T_1 \rangle$ slightly higher than $T_{\rm bath}$, as shown in Fig.~\ref{Fig5}b. This experiment was one of the first demonstrations of phase-coherent caloritronics, and proved that photonic heat conduction can reach the quantum limit when $Z_1=Z_2$ and $Z_{\rm c}$ is minimized~\cite{Meschke}.

One step beyond this experiment is based on the concept of a fully contactless photonic heat transistor~\cite{Ojanen,Pascal,Paolucci}, schematically represented in Fig.~\ref{Fig5}c. In this case, the N reservoirs (assumed to be made of Al$_{0.98}$Mn$_{0.02}$) are \textit{inductively} or \textit{capacitively} coupled to the intermediate circuit, which is based on a DC SQUID. The latter can be treated as a purely reactive LC circuit with a variable Josephson inductance. By heating the left reservoir up to $T_{\rm hot}$ yields flux-dependent $T_2$ modulations in the other remote N electrode, as displayed by the calculated curves in Fig.~\ref{Fig5}d. These theoretical curves have been obtained by assuming an inductive coupling with a mutual inductance of $500$ pH for different values of $T_{\rm hot}$ at $T_{\rm bath}=10$ mK and show remarkably large modulations up to 200 mK for $T_{\rm hot} =700$ mK and $\langle T_2 \rangle =170$ mK. Similar results can be obtained via a capacitive coupling, which turns out to be even more efficient and accessible from the fabrication point of view~\cite{Pascal,Paolucci}.

The photonic method opens the way to the investigation and exploitation of non-galvanic thermal transport, which would lead to wireless electronic cooling and to the remote control of noise and decoherence in mesoscopic quantum circuits. Although its effectiveness is maximized in the case of N leads, it has been predicted that the photonic coupling between S leads could still produce a significant thermal flow (two orders of magnitude higher than that produced by the electron-phonon coupling)~\cite{BosisioPRB2}. This possibility could lead to non-galvanic refrigerators for superconducting quantum circuits, in which a precise qubit initialization is required. For instance, a SINIS cooler~\cite{GiazottoRev,MuhonenRev} could be capacitively coupled by means of a DC SQUID to a part of the superconducting qubit structure. For the same purpose, a notable alternative approach has been experimentally demonstrated by exploiting photon assisted tunneling in a NIS junction, which can be used as a quantum circuit refrigerator~\cite{MottonenArx}.
It is also worth mentioning recent experimental results that proved quantum-limited heat conduction over macroscopic distances (up to a meter)~\cite{Partanen}, leading to even more possibilities for remote cooling.
Finally, the same kind of systems could also be the basis of photonic thermal rectifiers~\cite{Ruokola}.

\section*{Applications and future directions}

\begin{figure}
\centering
\includegraphics[width=1\columnwidth]{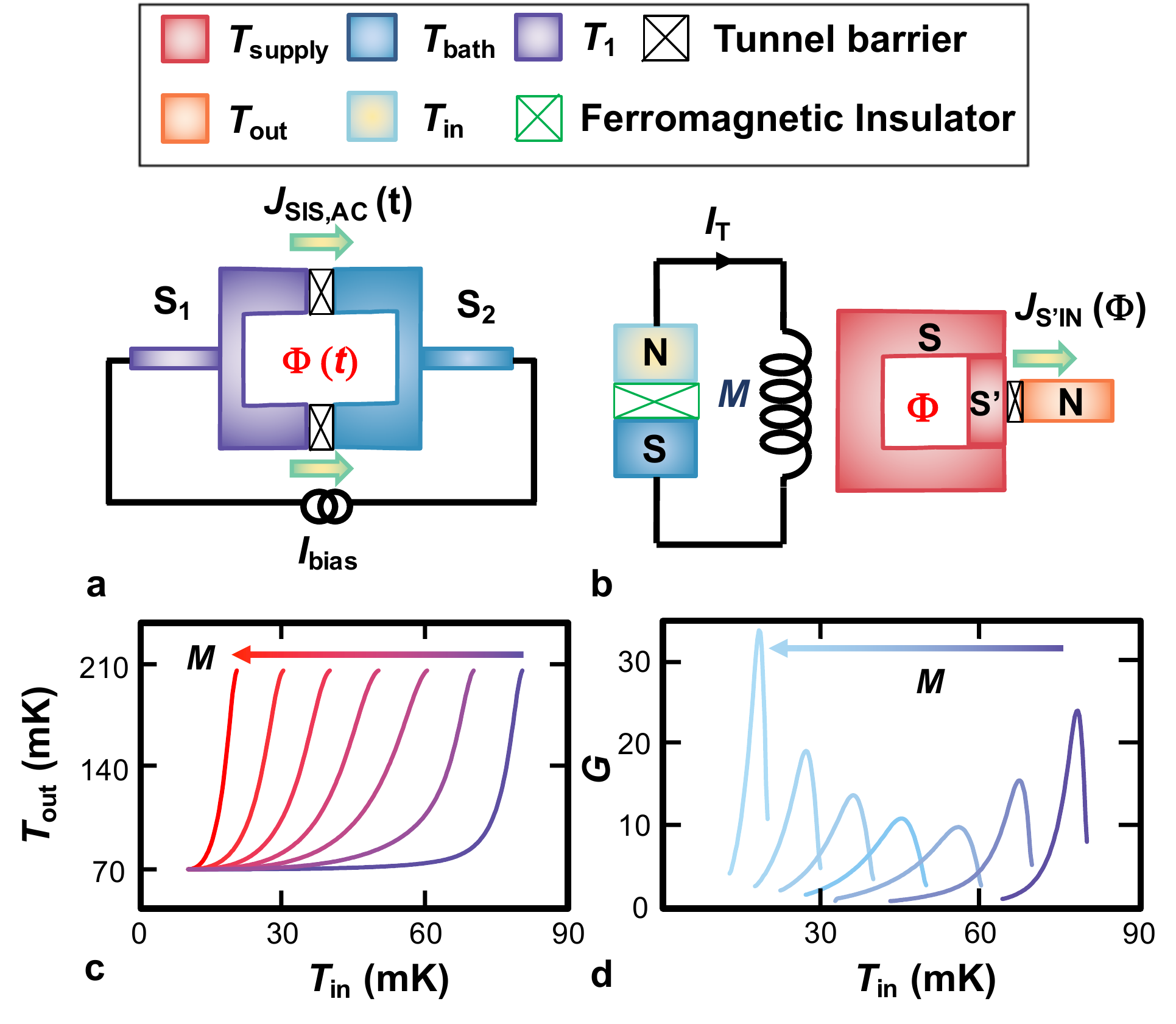}
\caption{\textbf{Future directions.} \textbf{a.} Schematic representation of a microwave Josephson refrigerator, which consists of a DC SQUID formed by two superconductors S$_1$ and S$_2$ biased by a small current $I_{\rm bias}$. The superconducting loop is pierced by a time-dependent magnetic flux $\Phi (t)$ that can induce a finite voltage bias across the device, which generates a heat current $J_{\rm SIS,AC}$ flowing through the JJs. The latter is used to cool the electronic temperature $T_1$ of S$_1$ below $T_{\rm bath}$. In order to obtain an efficient cooling power, it is necessary to have $\delta\equiv \Delta_2(0)/\Delta_1(0)>1$ and a finite capacitance of the JJs forming the SQUID to rectify $J_{\rm SIS,AC}$~\cite{SolinasPRB}. 
\textbf{b.} Sketch of a phase-coherent thermal amplifier based on the thermoelectric effect in a normal metal-ferromagnetic insulator-superconductor junction (assumed to be made of Cu-europium sulfide-Al). When the input temperature $T_{\rm in}$ is increased above $T_{\rm bath}$, a finite thermoelectric current can flow through a closed circuit including a superconducting coil. By means of the mutual inductance $M$, the latter generates a finite flux $\Phi$ that can control the thermal current flowing across a temperature-biased heat valve (presented in Fig.~\ref{Fig4}c). The S and S' parts of the valve are biased at $T_{\rm supply}$, while the N electrode resides at the output temperature of the amplifier $T_{\rm out}$~\cite{PaolucciAmpl}. \textbf{c.} Calculated $T_{\rm out}$ vs. $T_{\rm in}$ for different values of $M$ at $T_{\rm bath}=10$ mK and $T_{\rm supply}=250$ mK. \textbf{d.} Temperature differential gain $G=dT_{\rm out}/dT_{\rm in}$ as a function of $T_{\rm in}$ calculated for the curves shown in panel c. In panels c and d the arrows indicate increasing values of $M$.
\label{Fig6}} 
\end{figure}

All the structures discussed above can be implemented by conventional nanofabrication techniques, i.e., electron beam lithography, shadow mask evaporation of metals and \textit{in-situ} oxidation~\cite{Meschke,GiazottoNature,MartinezNature,MartinezNatRect,FornieriNature,FornieriArxiv}. They would join the well-known systems used for heating, cooling and thermometry in mesoscopic superconducting circuits~\cite{GiazottoRev}, which now embody a unique playground for investigating heat transport at the nanoscale. Moreover, this platform has already demonstrated that high levels of thermal isolation from the environment are achievable~\cite{Wei,Govenius}, showing significant room for improvement in the performance of caloritronic devices.

Phase-coherent caloritronics would offer many new possibilities to this field, as for instance the microwave Josephson refrigerator~\cite{SolinasPRB} shown in Fig.~\ref{Fig6}a. The latter consists of a DC SQUID made of two different superconductors S$_1$ and S$_2$ characterized by $\delta>1$ (indicating the thermal asymmetry in the system) and pierced by a time-dependent magnetic flux $\Phi(t)$. The latter induces phase oscillations across the JJs, which generate a finite voltage bias across the device and leads to the active cooling of S$_1$. The simplicity and the scalability of this system make it attractive for many superconducting quantum circuits that could be cooled at distance, thanks to high frequency modulations of $\Phi(t)$. Together with thermal rectifiers and non-galvanic photonic refrigerators, this system could be useful to evacuate unwanted hot quasiparticles from qubit architectures in preparation for quantum operations, thus improving their performance against decoherence.
Beyond quantum information~\cite{NielsenChuang}, phase-coherent caloritronics would certainly benefit many fields of nanoscience requiring an accurate administration of heat, including solid-state cooling~\cite{GiazottoRev}, energy harvesting, thermal isolation and radiation detection~\cite{GiazottoRev}. The advent of heat transistors and thermal memories could also pave the way to a new field called thermal logic~\cite{LiRev}, in which information is transferred, processed and stored under the form of energy. The latter could represent one of the most intriguing opportunities to exploit the (otherwise wasted) power dissipated by electronic circuits. In this context, another interesting proposal is an Al DC SQUID with a non-negligible inductance $L$. For significant values of $L$, the temperature interference pattern generated by the SQUID should exhibit a remarkable hysteresis, that could be used to obtain a thermal memory device~\cite{Guarcello,Guarcello2}.

Furthermore, caloritronic superconducting circuits might be combined with hybrid platforms based on, e.g., semiconducting nanowires~\cite{Mourik,Mastomaki}, graphene~\cite{Yokohama,Rainis,Paolucci}, ferromagnets or ferromagnetic insulators (FI)~\cite{Kawabata,GiazottoBergeret1,GiazottoBergeret2,GiazottoBergeret3}, topological insulators~\cite{Zhang,Ren1,Ren2,Sothmann1,Sothmann2} or low dimensional electronic devices to enhance their functionalities. For instance, Ref.~\citenum{Paolucci} theoretically analyzes a photonic heat transistor based on graphene reservoirs, whose carrier densities can be independently tuned thanks to two electrostatic gates. The latter can therefore control the photonic transfer function and the electron-phonon coupling in each reservoir, leading to increased temperature modulation amplitudes~\cite{Paolucci}. 

Even more interesting, Fig.~\ref{Fig6}b shows how the combination of a N-FI-S junction (assumed to be made of Cu-europium sulfide-Al) and a proximity heat valve, as the one discussed previously, can be used to design a very efficient temperature amplifier~\cite{PaolucciAmpl}. When the input temperature $T_{\rm in}$ of the N electrode is increased above $T_{\rm bath}$, a finite thermoelectric current $I_{\rm T}$ can flow through a closed circuit including a superconducting coil. By means of the mutual inductance $M$, the latter generates a finite flux $\Phi$ that can control the thermal current flowing across a temperature-biased heat valve (presented in Fig.~\ref{Fig4}c). The S and S' parts of the valve are biased at $T_{\rm supply}>T_{\rm bath}$, while the N electrode resides at the output temperature of the amplifier $T_{\rm out}$. Fig.~\ref{Fig6}c displays the behavior of $T_{\rm out}$ as a function of $T_{\rm in}$ for different values of $M$ between the coil and the heat valve at $T_{\rm bath}=10$ mK and $T_{\rm supply}=250$ mK. While the minimum and the maximum values of $T_{\rm out}$ are determined by $T_{\rm bath}$ and $T_{\rm supply}$, the value of $T_{\rm in}$ corresponding to the maximum output temperature decreases as the inductive coupling is raised. The temperature differential gain $G=dT_{\rm out}/dT_{\rm in}$ is shown in Fig.~\ref{Fig6}d and exceeds 10 in a large range of parameters. Although appearing more challenging from a fabrication point of view, this device can provide output temperatures in the same range as its input temperatures, thus representing a crucial element for the realization of thermal logic gates~\cite{LiRev}.

Finally, phase-coherent caloritronics can shed light on several fundamental energy- and heat-related phenomena at the nanoscale, such as quantum thermodynamics~\cite{PekolaNat}, heat transport in topological states of matter~\cite{Zhang,Ren1,Ren2,Sothmann1,Sothmann2} and give rise to phase-coherent thermoelectric effects in superconducting hybrid circuits~\cite{HeikkilaGiazotto,GiazottoMoodera,GiazottoDubi}.

\section*{Acknowledgments}
We thank J. P. Pekola and M. Meschke for providing experimental data. We also thank F. Paolucci, G. Timossi, E. Strambini and L. Casparis for fruitful discussions.
The MIUR-FIRB2013–Project Coca (grant no. RBFR1379UX) and the European Research Council under the European Union’s Seventh Framework Programme (FP7/2007-2013)/ERC grant agreement no. 615187 - COMANCHE are acknowledged for partial financial support.


\end{document}